\documentclass[aps,pra,floatfix,showkeys,twocolumn,10pt,citeautoscript,longbibliography]{revtex4-1}
\usepackage{amssymb}
\usepackage{amsmath}
\usepackage[pdftex]{graphicx}
\usepackage{dcolumn}
\usepackage{bm}
\usepackage{textcomp}
\usepackage[svgnames]{xcolor}
\usepackage{sidecap}

\usepackage[svgnames]{xcolor}
\definecolor{trueblue}{rgb}{0.0, 0.45, 0.81}
\definecolor{crimsonglory}{rgb}{0.75, 0.0, 0.2}
\definecolor{forestgreen}{rgb}{0.13, 0.55, 0.13}

\usepackage{hyperref}
\hypersetup{colorlinks,linkcolor={forestgreen},citecolor={crimsonglory},urlcolor={crimsonglory}}

\begin{document}

\title{Unveiling optical in-plane anisotropy of 2D materials from oblique incidence of light}

\author{M. Oliva-Leyva}
\email{mauriceoliva.cu@gmail.com}
\author{G. Gonzalez de la Cruz}
\email{bato@fis.cinvestav.mx}

\affiliation{Departamento de F\'isica, Centro de Investigaci\'on y de Estudios Avanzados del IPN, Apartado Postal 14-740, C.P. 07000, Ciudad de M\'exico, Mexico.}


\begin{abstract}
In this work, we present a theoretical study of the dispersion of linearly polarized light between two dielectric media separated by an anisotropic two-dimensional (2D) material under oblique incidence. Assuming that the 2D material is a conducting sheet of negligible thickness, generalized Fresnel coefficients are derived as a function of usual quantities (e.g. refraction indexes and scattering angles) and the anisotropic in-plane optical conductivity of the interstitial 2D material. In particular, we analyzed the modifications due to the 2D material of two classical optical phenomena: the Brewster effect and the total internal reflection. As an application, our general findings are particularized for uniaxially strained graphene. Effects of a uniaxial strain on the Brewster angle and the reflectance (under total internal reflection) are evaluated as a function of the magnitude and direction of strain.  
\vspace{2cm}
\end{abstract}

\maketitle

\section{Introduction}

Among the exceptional properties exhibited by the 2D materials, one can cite the strong interaction with light in comparison with their bulk counterparts \cite{Maier2013,Xia2014,Mak2016,Yakobson2018}. For example, graphene presents a constant and universal absorbance of $2.3\%$ across the whole infrared to visible spectral range \cite{Nair08,Mak2008}. On the other hand, monolayer  transition-metal  dichalcogenides, such as MoS$_2$ and WSe$_{2}$, are able to absorb over 10\% of incident light at the bandgap resonances \cite{Mak2010,Heinz2014}. In consequence, the incorporation of 2D materials in optical systems modifies their electrodynamical response. These modifications are observable in fundamental optical phenomena such as the Brewster effect \cite{Lin2016,Lambin2018,Pickwell2018}, the total internal reflection \cite{Tian2013,Pickwell2016,Pickwell2017} or the Goos--H\"{a}nchen shift \cite{Fan2016,You2018,Zambale2019}.

Another of the fascinating properties of the 2D materials is the high stretchability. They are capable of withstanding much larger elastic deformation compared to conventional electronic materials \cite{Akinwande2017}. For instance, graphene endures reversible stretching beyond of $10\%$ \cite{Lee08,Perez2014}. This long interval of elastic response results in substantial changes of the electronic, thermal, chemical and optical properties  \cite{Roldan2015,Naumis2017,Zhang2019}. Therefore, the mechanical strain has been widely proposed as a tool to tune the physical properties of 2D materials and to ultimately achieve high-performance 2D-material-based devices. 

In particular, strain engineering of the optical response of graphene has been experimentally archived in various works \cite{Pereira14,Tian2014,Chhikara2017}. When graphene is under uniform strain, its electronic structure is modified, e.g. the Dirac cones change of shape, from circular to elliptical cross-section. As a consequence, the Fermi velocity becomes anisotropic \cite{Oliva2013,Oliva2017}. This strain-induced effect on the Fermi velocity is traduced in an anisotropic optical conductivity \cite{Pereira10,Pellegrino11,Oliva2014} which yields a modulation of the  transmittance for normal incidence of linearly polarized light as a function of the polarization direction \cite{Pereira10,Oliva2015}. However, to the best of our knowledge, the problem of light scattering for oblique incidence in the case of strained graphene (or an anisotropic 2D material) has not been analyzed in details. For instance, the contribution of the non-diagonal components of the strained graphene conductivity tensor on the light scattering remains unexplored. Thus, the main objective of this paper is to give a general characterization of this problem, which would lead a more complete understanding of the Brewster effect or the total internal reflection as a function of strain.

This paper is organized as follows. In Sec.~\ref{II}, we derive the generalized Fresnel coefficients when two dielectric media are separated by an anisotropic 2D material. Section \ref{III} is devoted to analyze the general features of the modifications of the Brewster effect or the total internal reflection due the optical anisotropy of the 2D material. In Sec. \ref{IV}, our findings are particularized to the case of graphene under a uniaxial strain deformation and we discuss previous experiments about total internal reflection \cite{Tian2014}. Finally, in Sec. \ref{V}, our conclusions are given.

\section{Generalized Fresnel coefficients}\label{II}

\begin{figure}[t]
\includegraphics[width=\linewidth]{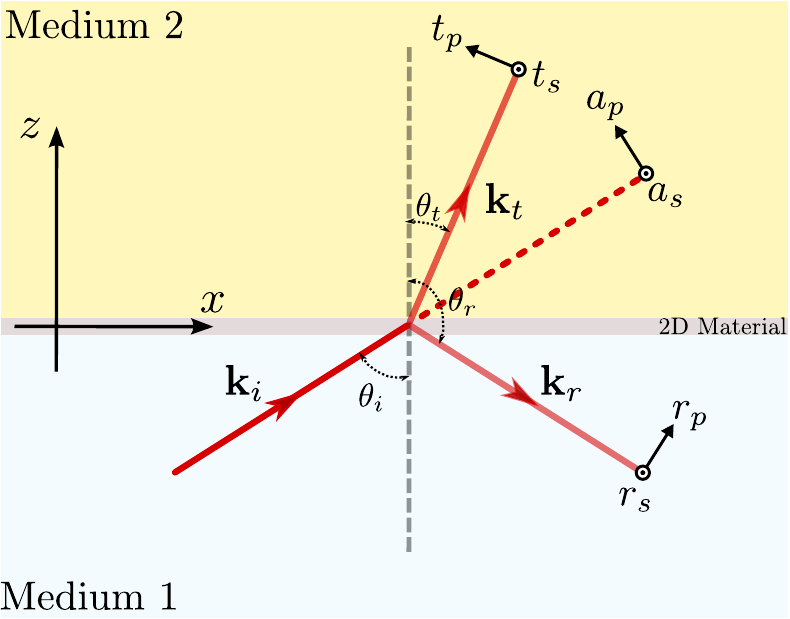}
\caption{Front view of the scattering geometry for oblique incidence between two media with a 2D conducting material separating them.}
\label{fig1}
\end{figure}

We consider the propagation of linearly polarized light through two semi-infinite non-absorbing dielectric media separated by a 2D material and with refractive indices $n_{1}$ and $n_{2}$, as shown schematically in Fig. \ref{fig1}. In general, the 2D material is assumed to be an anisotropic conducting sheet of negligible thickness \cite{Merano2016,Heinz2018}, whose optical conductivity is characterized by the second-rank symmetric tensor $\overline{\bm{\sigma}}$. According to this model, the effects of the out-plane anisotropy, recently investigated by Maj\'erus \emph{et.al.}\cite{Majerus2018}, are disregarded. It is worth mentioning that the optical in-plain anisotropy of the 2D material can be an intrinsic property, as occurred for example in phosphorene \cite{Wang2016} and borophene \cite{Lherbier2016}, or induced by strain \cite{Nguye2017}.

For the considered scattering problem, the components of the electric ($\bm{E}$) and magnetic ($\bm{H}$) fields in the incident, reflected and transmitted waves can be written as~\cite{Born}:

\emph{incident wave}
\begin{align}
E_{x}^{(i)}&=-a_{p}\cos\theta_{i} e^{i\tau_{i}}, &  H_{x}^{(i)}z_{0}&=-a_{s}\cos\theta_{i}n_{1}e^{i\tau_{i}}, \nonumber \\ 
E_{y}^{(i)}&=a_{s} e^{i\tau_{i}}, & H_{y}^{(i)}z_{0}&=-a_{p}n_{1} e^{i\tau_{i}}, \nonumber\\ 
E_{z}^{(i)}&=a_{p}\sin\theta_{i} e^{i\tau_{i}}, & 
H_{z}^{(i)}z_{0}&=a_{s}\sin\theta_{i}n_{1} e^{i\tau_{i}},\label{FI}
\end{align}

\emph{reflected wave}
\begin{align}
E_{x}^{(r)}&=-r_{p}\cos\theta_{r} e^{i\tau_{r}}, &  H_{x}^{(r)}z_{0}&=-r_{s}\cos\theta_{r}n_{1} e^{i\tau_{r}}, \nonumber \\ 
E_{y}^{(r)}&=r_{s} e^{i\tau_{r}}, & 
H_{y}^{(r)}z_{0}&=-r_{p}n_{1} e^{i\tau_{r}}, \nonumber\\ 
E_{z}^{(r)}&=r_{p}\sin\theta_{r} e^{i\tau_{r}}, & 
H_{z}^{(r)}z_{0}&=r_{s}\sin\theta_{r}n_{1} e^{i\tau_{r}},\label{FR}
\end{align}

and \emph{transmitted wave}
\begin{align}
E_{x}^{(t)}&=-t_{p}\cos\theta_{t} e^{i\tau_{t}}, &  H_{x}^{(t)}z_{0}&=-t_{s}\cos\theta_{t}n_{2} e^{i\tau_{t}}, \nonumber \\ 
E_{y}^{(t)}&=t_{s} e^{i\tau_{t}}, & H_{y}^{(t)}z_{0}&=-t_{p}n_{2} e^{i\tau_{t}}, \nonumber\\ 
E_{z}^{(t)}&=t_{p}\sin\theta_{t} e^{i\tau_{t}}, & 
H_{z}^{(t)}z_{0}&=t_{s}\sin\theta_{t}n_{2} e^{i\tau_{t}},\label{FT}
\end{align}
where $z_{0}$ is the vacuum impedance and $\tau_{j}=\bm{k}_{j}\cdot\bm{r}-\omega t$ ($j=i,r,t$), with $\omega$ being the angular frequency of light and $\bm{k}_{j}$  the respective wave vector. The relation between $\bm{E}$ and $\bm{H}$ for each wave is $\bm{H}z_{0}=n(\bm{k}/\vert\bm{k}\vert)\times\bm{E}.$

At the interface $z=0$, the electric and magnetic fields are related by the boundary conditions \cite{Jackson},
\begin{align}
E_{x}^{(t)}-E_{x}^{(i)}-E_{x}^{(r)}&=0, & E_{y}^{(t)}-E_{y}^{(i)}-E_{y}^{(r)}&=0,\label{BCE}\\
H_{x}^{(t)}-H_{x}^{(i)}-H_{x}^{(r)}&=J_{y}, & H_{y}^{(t)}-H_{y}^{(i)}-H_{y}^{(r)}&=-J_{x},\label{BCH}
\end{align}
where $\bm{J}$ is the surface current density because of the 2D conducting material. According to the Ohm's law $\bm{J}=\overline{\bm{\sigma}}\cdot\bm{E}^{(t)}$, namely
\begin{align}
J_{x}&=\sigma_{xx}E_{x}^{(t)}+\sigma_{xy}E_{y}^{(t)}, \nonumber\\ J_{y}&=\sigma_{yy}E_{y}^{(t)}+\sigma_{xy}E_{x}^{(t)},\label{J}
\end{align}
with $\sigma_{xx}$, $\sigma_{yy}$ and $\sigma_{xy}$ being the components of the optical conductivity tensor $\overline{\bm{\sigma}}$. To note that $\overline{\bm{\sigma}}$ is symmetric, i.e. $\sigma_{xy}=\sigma_{yx}$.

Now, substituting Eqs.~(\ref{FI}-\ref{FT}) into Eqs.~(\ref{BCE}-\ref{J}), and using the fact that $\cos\theta_{r}=-\cos\theta_{i}$, we obtain the components of the reflected and transmitted waves, in terms of those of the incident wave, are given through the following generalized Fresnel coefficients,
\begin{widetext}
\begin{align}
r_{s}&=\Bigl(-1 + \frac{2 n_{1}f_{1}\cos\theta_{i}}{f_{1}f_{2}+z_{0}^{2}\sigma_{xy}^{2}\cos\theta_{i}\cos\theta_{t}}\Bigr) a_{s} + \Bigl(\frac{2 n_{1}z_{0}\sigma_{xy}\cos\theta_{i}}{f_{1}f_{2}+z_{0}^{2}\sigma_{xy}^{2}\cos\theta_{i}\cos\theta_{t}}\Bigr) a_{p}, \label{Rs}\\
t_{s}&=\Bigl(\frac{2 n_{1}f_{1}\cos\theta_{i}}{f_{1}f_{2}+z_{0}^{2}\sigma_{xy}^{2}\cos\theta_{i}\cos\theta_{t}}\Bigr) a_{s} + \Bigl(\frac{2 n_{1}z_{0}\sigma_{xy}\cos\theta_{i}}{f_{1}f_{2}+z_{0}^{2}\sigma_{xy}^{2}\cos\theta_{i}\cos\theta_{t}}\Bigr) a_{p}, \\
r_{p}&=\Bigl(1 - \frac{2 n_{1}f_{2}\cos\theta_{t}}{f_{1}f_{2}+z_{0}^{2}\sigma_{xy}^{2}\cos\theta_{i}\cos\theta_{t}}\Bigr)a_{p} + \Bigl(\frac{2 n_{1}z_{0}\sigma_{xy}\cos\theta_{t}}{f_{1}f_{2}+z_{0}^{2}\sigma_{xy}^{2}\cos\theta_{i}\cos\theta_{t}}\Bigr)a_{s}, \\
t_{p}&=\Bigl(\frac{2 n_{1}f_{2}\cos\theta_{i}}{f_{1}f_{2}+z_{0}^{2}\sigma_{xy}^{2}\cos\theta_{i}\cos\theta_{t}}\Bigr) a_{p} - \Bigl(\frac{2 n_{1}z_{0}\sigma_{xy}\cos\theta_{i}}{f_{1}f_{2}+z_{0}^{2}\sigma_{xy}^{2}\cos\theta_{i}\cos\theta_{t}}\Bigr) a_{s},\label{Tp} 
\end{align} 
with $f_{1}=(n_{1}\cos\theta_{t}+n_{2}\cos\theta_{i}+z_{0}\sigma_{xx}\cos\theta_{i}\cos\theta_{t})$ and $f_{2}=(n_{1}\cos\theta_{i}+n_{2}\cos\theta_{t}+z_{0}\sigma_{yy})$.
\end{widetext}

\begin{figure*}[t]
\includegraphics[width=\linewidth]{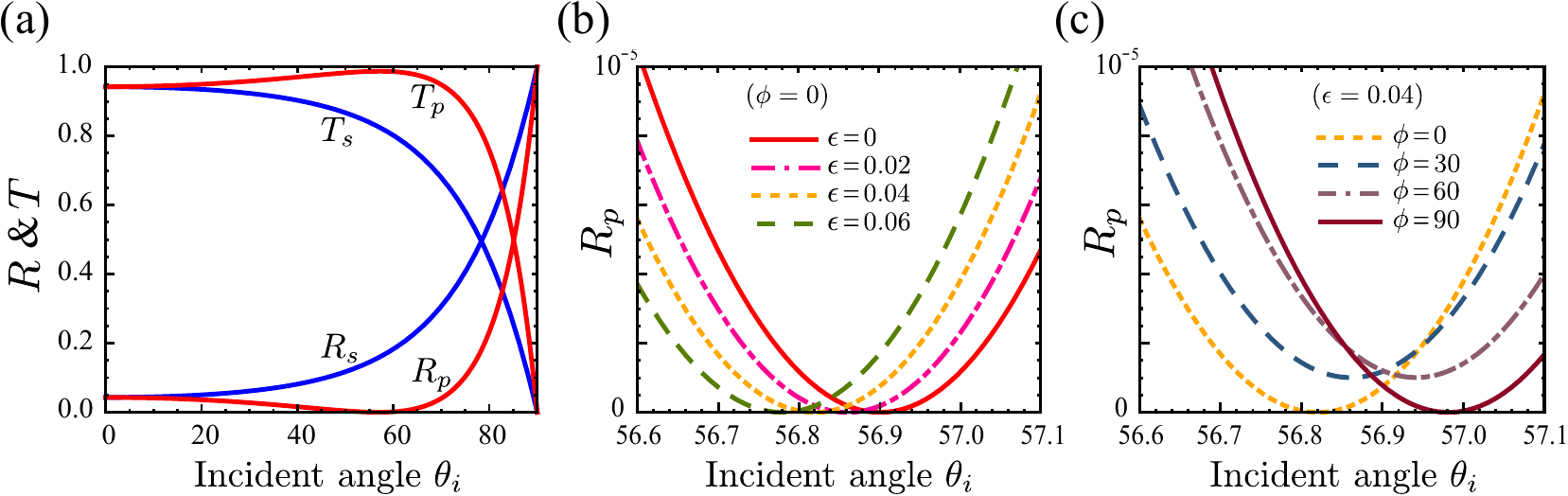}
\caption{(a) Reflectance and transmittance of $s-$ and $p$-polarized waves as functions of the incident angle $\theta_{i}$ for a system dielectric-graphene-dielectric with $n_{1}=1$ and $n_{2}=1.5$. Graphene is assumed to be unstrained and with optical conductivity equals to $\pi\alpha/z_{0}$. (b)
Reflectance of $p$-polarized radiation around the Brewster angle of a system as (a), but now graphene is uniaxially stretched along of the $x-$axis under different strain magnitudes. (c) Analogous to panel (b), but for different directions of a uniaxial strain of magnitude $\epsilon=0.04$.}
\label{fig2}
\end{figure*}

\section{Results}\label{III}

As a first result from these equations one can note that, in general, the considered scattering of light can not be decoupled in transverse electric (TE) and transverse magnetic (TM) waves. When the normal to the incidence plane is not along one principal axes of the conductivity tensor $\overline{\bm{\sigma}}$, both incident TE ($s$-polarized) and TM ($p$-polarized) waves are scattered in reflected and transmitted waves with components that are parallel and perpendicular to the incidence plane. Namely, an incident $s$- or $p$-polarization is only preserved if $\sigma_{xy}=0$. In this last case, the Fresnel coefficients (\ref{Rs}-\ref{Tp}) are quite more simplified and, in particular, the reflectance and transmittance for both $s$- and $p$-polarization can be calculated as 
\begin{align}
R_{s}&=\Bigl\vert\frac{n_{1}\cos\theta_{i}-n_{2}\cos\theta_{t}-z_{0}\sigma_{yy}}{n_{1}\cos\theta_{i}+n_{2}\cos\theta_{t}+z_{0}\sigma_{yy}}\Bigr\vert^{2}, \label{R-s}\\
R_{p}&=\Bigl\vert\frac{n_{2}/\cos\theta_{t}-n_{1}/\cos\theta_{i}+z_{0}\sigma_{xx}}{n_{2}/\cos\theta_{t}+n_{1}/\cos\theta_{i}+z_{0}\sigma_{xx}}\Bigr\vert^{2},\label{R-p} \\
T_{s}&=\frac{4n_{1}n_{2}\cos\theta_{i}\cos\theta_{t}}{\big\vert n_{1}\cos\theta_{i}+n_{2}\cos\theta_{t}+z_{0}\sigma_{yy}\big\vert^{2}} , \\
T_{p}&=\frac{4n_{1}n_{2}/(\cos\theta_{i}\cos\theta_{t})}{\big\vert n_{2}/\cos\theta_{t}+n_{1}/\cos\theta_{i}+z_{0}\sigma_{xx}\big\vert^{2}} ,\label{T-p} 
\end{align}
and the absorbance as $A=1-R-T$.
These expressions reproduce those for an isotropic 2D material like unstrained graphene \cite{Zhan2013,Lambin2015} if both $\sigma_{xx}$ and $\sigma_{yy}$ are replacing with the same conductivity value. In absence of anisotropy and normal incidence, the reflectance, transmittance and absorbance are independent on the incident light polarization, as depicted in Fig.~\ref{fig2}(a), Fig.~\ref{fig3}(a) and  Fig.~\ref{fig4}(a), respectively. However, for normal incidence but in the presence of anisotropy, from Eqs.~(\ref{R-s}--\ref{T-p}) it is clear that they depend on the light polarization, which has been studied in details for strained graphene \cite{Oliva2015}.

\subsection{Modified Brewster angle}

\begin{figure*}[t]
\includegraphics[width=0.75\linewidth]{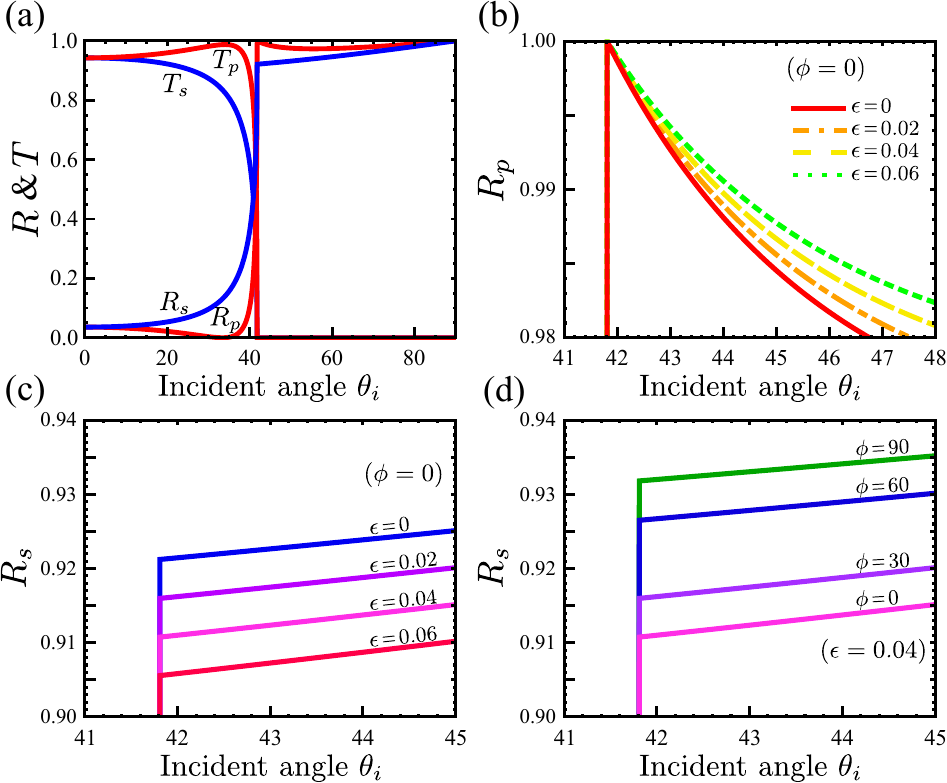}
\caption{(a) Reflectance and transmittance of $s$- and $p$-polarized waves as functions of the incident angle $\theta_{i}$ for a system dielectric-graphene-dielectric with $n_{1}=1.5$ and $n_{2}=1$. Graphene is assumed to be unstrained and with optical conductivity equals to $\pi\alpha/z_{0}$. Panels (b) and (c), show respectively the reflectance of $p$- and $s$-polarized waves around the critical angle of a system as (a), but now graphene is uniaxially stretched along of the $x$-axis. (d) Analogous to panel (c), but for different directions of a uniaxial strain of magnitude $\epsilon=0.04$.}
\label{fig3}
\end{figure*}

For oblique incidence, such quantities show a strong dependence on the light polarization. For instance, while the $s-$polarized radiation is partially reflected for all $\theta_{i}$, the $p-$polarization reflection is totally inhibited at the Brewster angle (see Fig.~\ref{fig2}(a) and Fig.~\ref{fig3}(a)). For the simplest system dielectric-dielectric, the Brewster angle $\theta_{\text{B}}$ is given by the well-know formula $\tan\theta_{\text{B}}=n_{2}/n_{1}$ \cite{Born}. Once a 2D conducting material is present at the interface between both dielectric media, this angle changes \cite{Lambin2018}. 

Staring from the cancellation condition of $R_{p}$ given by Eq.~(\ref{R-p}) and using the Snell law, the modified Brewster angle $\theta_{\text{B}}^{\prime}$ can be expressed as 
\begin{equation}
\theta_{\text{B}}^{\prime}\approx\theta_{\text{B}}+\frac{z_{0}\sigma_{xx}n_{1}n_{2}^{3}}{(n_{2}^{2} - n_{1}^{2})(n_{2}^{2}+n_{1}^{2})^{3/2}},\label{BA}
\end{equation}
which is fulfilled in the regime of purely real and low conductivity, i.e. $\Im(\sigma_{xx})=0$ and $z_{0}\sigma_{xx}\ll1$. Therefore, the shift of the Brewster angle $\Delta_{\text{B}}=\theta_{\text{B}}^{\prime}-\theta_{\text{B}}$ is imposed by the relation between the refractive indexes of the incident medium $n_{1}$ and of the substrate $n_{2}$: \emph{if} $n_{2}>n_{1}$ ($n_{1}>n_{2}$) \emph{then} $\Delta>0$ ($\Delta<0$). Equation~(\ref{BA}) also leads to useful approximated expression for the following limiting cases:
\begin{align}
&\text{if}\ \ n_{2} \gg n_{1}, & \Delta_{\text{B}}&\approx z_{0}\sigma_{xx}n_{1}/n_{2}^{2}, \label{AAB}\\
&\text{if}\ \ n_{2} \ll n_{1}, & \Delta_{\text{B}}&\approx-z_{0}\sigma_{xx}n_{2}^{3}/n_{1}^{4}, 
\end{align}
being the approximation (\ref{AAB}) coincident with the reported one in Ref.~[\onlinecite{Lambin2018}]. 

\subsection{``Total'' internal reflection}

Another fundamental optical effect occurs when the incident medium is optically denser than the substrate. For such system (with $n_{1}>n_{2}$), if the incident angle exceeds the critical value $\theta_{c}$ given by $\sin\theta_{c}=n_{2}/n_{1}$ then all the incident light is reflected, i.e. $R_{s,p}=1$ and $T_{s,p}=0$ for $\theta_{i}\geq\theta_{c}$, which is know as total internal reflection \cite{Born}. The presence of a 2D conducting material at the interface modifies this phenomenon \cite{Tian2013,Tian2014,Zhan2013}. Unlike the Brewster angle, the critical angle $\theta_{c}$ does not change. However, while the transmittance remains nullified for $\theta_{i}\geq\theta_{c}$, the reflectance is no longer total, as it can be appreciated in Fig.~\ref{fig3}(a). Above the critical angle, the $p$-polarization reflectance $R_{p}$ roughly presents a very slow U-shaped behaviour (concave downward) as a function of the incident angle $\theta_{i}$, being equal to $1$ for $\theta_{i}=\theta_{c}$ and $\theta_{i}=90\text{\textdegree}$. Such U-shaped behaviour of $R_{p}$ is more pronounced for a 2D material with higher optical conductivity $\sigma_{xx}$, whereas $R_{p}$ recovers its value $1$ for $\sigma_{xx}$ tending to $0$. This fact simply means that under total internal reflection the $p$-polarization  absorbance $A_{p}=1-R_{p}$ increases as $\sigma_{xx}$ increases.

On the other hand, for $\theta_{i}\geq\theta_{c}$ the $s$-polarization reflectance $R_{s}$ shows an approximate (increasing) lineal dependence of $\theta_{i}$, reaching to maximum value $1$ at  $\theta_{i}=90\text{\textdegree}$. Using Eq.~(\ref{R-s}) one can be derived that $R_{s}$ at the critical angle $\theta_{c}$ takes the exact value:
\begin{equation}\label{RsAc}
\overline{R}_{s}\equiv {R}_{s}(\theta_{c})=\Bigl\vert\frac{\sqrt{n_{1}^{2}-n_{2}^{2}}-z_{0}\sigma_{yy}}{\sqrt{n_{1}^{2}-n_{2}^{2}}+z_{0}\sigma_{yy}}\Bigr\vert^{2},   
\end{equation}
which could be used to determine the optical conductivity of 2D materials from reflectance measurements of $s$-polarized radiation under configuration of total internal reflection. It is important to note that the $s$-polarization absorbance $A_{s}$ has its maximum at $\theta_{c}$, as illustrated in Fig.~\ref{fig4}(a), being equal to $1-\overline{R}_{s}$. Then, from Eq.~(\ref{RsAc}) and neglecting the higher-order term of $\sigma_{yy}$, the maximum of $A_{s}$ is given by 
\begin{equation}
\overline{A}_{s}\equiv {A}_{s}(\theta_{c})\approx\frac{4z_{0}\Re(\sigma_{yy})}{\sqrt{n_{1}^{2} - n_{2}^{2}}}.
\end{equation}
Thus, an increasing of the optical conductivity $\sigma_{yy}$ leads to an increasing of the $s$-polarization absorbance peak (see Fig.~\ref{fig4}(c)).

\section{Strained graphene}\label{IV}

In order to apply our previous results to the concrete anisotropic 2D material, let us consider graphene under uniaxial strain. In general, when graphene is subjected to an arbitrary uniform strain (e.g. uniaxial, biaxial, and so forth) its optical conductivity tensor $\overline{\bm{\sigma}}$, up to first order in the strain tensor $\overline{\bm{\epsilon}}$, can be written as \cite{Pereira10,Pellegrino11,Oliva2014}
\begin{align}
\sigma_{xx}&=\sigma_{0}(1+\beta(\epsilon_{yy}-\epsilon_{xx})), \nonumber\\ \sigma_{yy}&=\sigma_{0}(1+\beta(\epsilon_{xx}-\epsilon_{yy})), \nonumber\\
\sigma_{xy}&=\sigma_{yx}=-2\beta\epsilon_{xy}\sigma_{0},\label{Cond}
\end{align}
where $\sigma_{0}$ is the conductivity of unstrained graphene and $\beta\simeq2.37$ is a parameter related to the hopping changes due to strain \cite{Ribeiro2009,Botello2018}. For chemical potential equals to zero and at near-infrared and visible frequencies, $\sigma_{0}$ takes the universal value $\pi\alpha/z_{0}$, where $\alpha\approx1/137$ is the fine structure constant.

In the case of a uniaxial strain such that the stretching direction is rotated by an angle $\phi$ respect to the $x$-axis of the laboratory frame, then the strain tensor components read
\begin{align}
\epsilon_{xx}&=\epsilon(\cos^{2}\phi - \nu\sin^{2}\phi), \nonumber\\ \epsilon_{yy}&=\epsilon(\sin^{2}\phi - \nu\cos^{2}\phi), \nonumber\\
\epsilon_{xy}&=\epsilon_{yx}=\epsilon(1+\nu)\sin\phi\cos\phi,
\end{align}
where $\epsilon$ is the strain magnitude and $\nu\approx0.16$ is the Poisson ratio. In consequence, from Eqs.~(\ref{Cond}) it follows that $\sigma_{xx,yy}=\sigma_{0}(1\mp\beta\epsilon(1+\nu)\cos2\phi)$ meanwhile $\sigma_{xy}=-\sigma_{0}\beta\epsilon(1+\nu)\sin2\phi$. So, for a stretching along the $x-$axis ($\phi=0\text{\textdegree}$) or the $y-$axis ($\phi=90\text{\textdegree}$), the conductivity component $\sigma_{xy}$ results zero and, therefore, the reflectance and transmittance can obtained from Eqs.~(\ref{R-s}-\ref{T-p}). Moreover, it is important to keep in mind that, according to Eqs.~(\ref{Cond}), the optical conductivity parallel (perpendicular) to the stretching direction decreases (increases) linearly with increasing the strain magnitude $\epsilon$, which has been experimentally observed \cite{Pereira14,Chhikara2017}.

\begin{figure*}[t]
\includegraphics[width=0.75\linewidth]{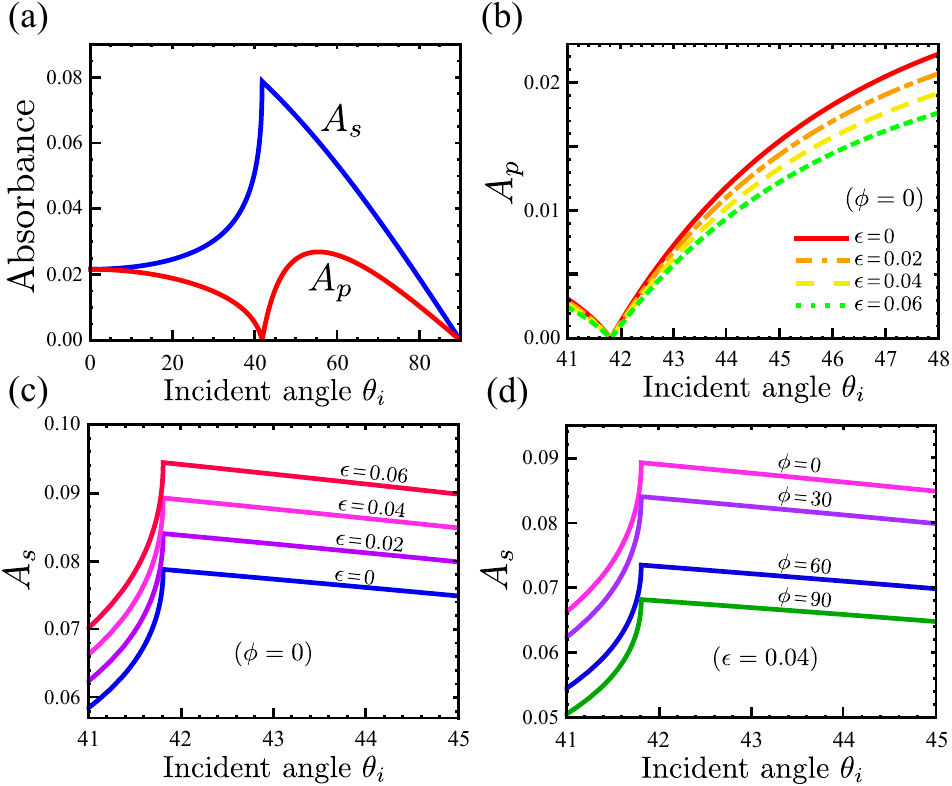}
\caption{(a) Dependence of the $s$- and $p$-polarization absorbance on the incident angle $\theta_{i}$ for $n_{1}=1.5$, $n_{2}=1$ and the graphene conductivity given by $\pi\alpha/z_{0}$. Panels (b) and (c), display respectively the $s$- and $p$-polarization absorbance around the critical angle for graphene uniaxially stretched along of the $x$-axis. (d) Similar to panel (c) but for different directions of a uniaxial strain of magnitude $\epsilon=0.04$.}
\label{fig4}
\end{figure*}

Figure~\ref{fig2}(b) depicts the $p$-polarization reflectance $R_{p}$ about the modified Brewster angle for uniaxially strained graphene along the $x-$axis under different strain magnitudes. The observed shift of $\theta_{\text{B}}^{\prime}$ for stronger stretching can be understood from Eq.~(\ref{AAB}). Replacing $\sigma_{xx}$ by its corresponding value, one gets
\begin{equation}
\Delta_{\text{B}}(\epsilon)\approx z_{0}\sigma_{0}(1-\beta\epsilon(1+\nu)) n_{1}/n_{2}^{2}.\label{AABs}
\end{equation}

Expression~(\ref{AABs}) allows easily to evaluate the modified Brewster angle as function on the strain magnitude $\epsilon$. For instance, for $n_{1}=1$, $n_{2}=1.5$ and $z_{0}\sigma_{0}=\pi\alpha$, it predicts a decreasing of $-0.03\text{\textdegree}$ for each $1\%$ that $\epsilon$ increases, with a relative error not greater than $2\%$ (see TABLE~\ref{table} and also Ref.~[\onlinecite{Lambin2018}]). Inversely, it is important to note that Eq.~(\ref{AABs}) could be used to determine the strain magnitude from measurements of the Brewster angle shift. 

\begin{table}[h]
\begin{tabular}{ |c|c|c| } 
 \hline
 $\epsilon$ & Exact $\theta_{\text{B}}^{\prime}$ & Approx. $\theta_{\text{B}}^{\prime}$ \\ 
  \hline
 0.00 & 56.90\textdegree & 56.89\textdegree \\ 
  \hline
 0.02 & 56.86\textdegree & 56.85\textdegree \\
 \hline
 0.04 & 56.82\textdegree & 56.81\textdegree \\
 \hline
 0.06 & 56.78\textdegree & 56.77\textdegree \\
 \hline
\end{tabular}
\caption{Modified Brewster angle for $n_{1}=1$, $n_{2}=1.5$, $z_{0}\sigma_{0}=\pi\alpha$ and different strain magnitudes (as in Fig.~\ref{fig2}(b)). The approximate values were obtained using Eq.~(\ref{AABs}).}
\label{table}
\end{table}

On the other hand, Fig.~\ref{fig2}(c) illustrates $R_{p}$ for uniaxial strains with the same magnitude ($\epsilon=0.04$), but along of different directions. The most striking feature is that $R_{p}$ only nullify for the elongations along the $x$- and $y$-axes. This fact is due to the non-diagonal component of the graphene conductivity is not zero for uniaxial strain along any other direction ($\sigma_{xy}=-\sigma_{0}\beta\epsilon(1+\nu)\sin2\phi$). As mentioned above, under an incident $p$-polarized wave, if $\sigma_{xy}\neq0$ then the reflected wave is not strictly $p$-polarized since its electric field also presents a small transverse component to the incidence plane (see Eq.~(\ref{Rs})), which inhibits the perfect suppression of $R_{p}$. In short, the Brewster effect does not happens in strained graphene whenever one principal axes of the strain tensor is not normal to the incidence plane. 

Finally, we extend our previous general discussion on the total internal reflection phenomenon to the case that the interstitial 2D material is strained graphene. Figure~\ref{fig3}(b) shows $R_{p}$ about the critical angle $\theta_{c}$ for $n_{1}=1.5$, $n_{2}=1$ and graphene uniaxially strained along the $x$-axis. For $\theta_{i}\geq\theta_{c}$, it can be noted that $R_{p}$ increases for stronger elongation. As above mentioned, this behavior is expected since the strain increasing reduces the conductivity $\sigma_{xx}$ and, therefore, $R_{p}$ tends to $1$ which is its value in the absence of graphene. As a consequence, under these basic considerations the $p$-polarization absorbance $A_{p}$ should decreases with increasing $\epsilon$ (see Fig.~\ref{fig4}(b)). However, Dong and coauthors \cite{Tian2014} experimentally observed an opposite dependence of $A_{p}$ on the strain magnitude $\epsilon$. Then, according to the discussion presented here, those results can not be understood in terms of strain-induced effects on graphene conductivity. 

Otherwise, Fig.~\ref{fig3}(c) displays the step-shape of $R_{s}$ about the critical angle $\theta_{c}$, which experiments a decreasing (an approximate down-shift curve) for stronger uniaxial strains along the $x$-axis. From Eq.~(\ref{RsAc}) and considering that $\sigma_{yy}=\sigma_{0}(1+\beta\epsilon(1+\nu))$, the value of $R_{s}$ at the critical angle $\theta_{c}$ as a function of $\epsilon$ results 
\begin{equation}\label{Rs-SG}
\overline{R}_{s}(\epsilon)\approx 1-4 z_{0}\sigma_{0}(1+\beta\epsilon(1+\nu))/\sqrt{n_{1}^{2}-n_{2}^{2}},
\end{equation}
which predicts the $\overline{R}_{s}$ decreasing when $\epsilon$ increases. In particular, for $n_{1}=1.5$, $n_{2}=1$ and $z_{0}\sigma_{0}=\pi\alpha$, from this relation it follows that the $\overline{R}_{s}$ value reduces approximately $-0.003$ for each $1\%$ that $\epsilon$ enhances. This variation is also experimented by $R_{s}$ for incident angles slightly above the angle critical, as noted in Fig.~\ref{fig3}(c). Therefore, the experimental monitoring of $R_{s}$ under total internal reflection conditions, as made in Ref.~[\onlinecite{Tian2014}], it would allow to investigate the strain-induced effects on graphene conductivity using Eq.~(\ref{Rs-SG}) or the most general expression~(\ref{RsAc}). It is worth mentioning that this study can be complemented by the consideration of ``the freedom degree'' associated to parameter $\phi$. Figure~\ref{fig3}(d) shows $R_{s}$ for uniaxial strains with $\epsilon=0.04$, but different directions. It can noted that $\overline{R}_{s}$ have a minimum when the elongation is parallel to the incidence plane, e.g. $\overline{R}_{s}\approx0.911$ for $\phi=0\text{\textdegree}$, while $\overline{R}_{s}\approx0.932$ for $\phi=90\text{\textdegree}$. This means an absorbance variation greater than $2\%$ due exclusively to the orientation change of the incidence plane respect to the strain direction, as appreciated in Fig.~\ref{fig4}(d). 

 \section{Conclusion}\label{V}
 
In summary, we derived the Fresnel coefficients for oblique incidence of linearly polarized light through two dielectric media with an anisotropic 2D conducting material at the interface. Based on these generalized coefficients it was demonstrated that, if the incidence plane is not parallel to the principal axes of the optical conductivity tensor of the 2D material, then the light scattering problem can not be decoupled in pure $s$- and $p$-polarized waves. In other words, whenever the non-diagonal conductivity component $\sigma_{xy}\neq0$ is not zero, the incident polarization is not preserved because the polarization plane changes. We performed an analysis about the modifications of the Brewster effect and the total internal reflection due to the anisotropic 2D material. For the former, an analytical expression of the modified Brewster angle has been obtained for low conductivity, which predicts a up-shift (down-shift) of the Brewster angle if the refractive index of the substrate is greater (smaller) than the one of the incident medium. This expression also reproduces a previous result found by Maj\'erus \emph{et.al.}\cite{Lambin2018} in certain limiting case. Moreover, it is demonstrated that for $\sigma_{xy}\neq0$ the perfect suppression of the $p$-polarization reflectance is inhibited and, thus, the Brewster effect does not occur strictly. 

To exemplify our findings, as anisotropic 2D material we considered uniaxially strained graphene. The uniaxial-strain effects on the Brewster angle and the reflectance and absorbance under total internal reflection were estimated as a function either of the magnitude or of the strain direction. The presented results also suggest that measurements of the modified Brewster angle and the $s$-polarization reflectance about the angle critical could be used to investigate the strain-induced effects on the optical conductivity of graphene and, in general, the optical anisotropy of 2D materials.

\begin{acknowledgments}
This work is partially supported by Conacyt-Mexico under Grant No. 254414.
\end{acknowledgments}




\bibliography{biblio}

\end{document}